# Optimization of LANSCE Proton Beam Performance for Isotope Production

Yuri K. Batygin

Los Alamos National Laboratory, Los Alamos, NM 87545, USA

*Abstract*

The LANSCE Isotope Production Facility (IPF) utilizes a 100-MeV proton beam with average power of 23 kW for isotope production in the fields of medicine, nuclear physics, national security, environmental science and industry. Typical tolerable fractional beam loss in the 100-MeV beamline is 4 x10$^{-3}$. Experimental beam study was made to minimize the beam losses. Adjustments to the ion source's extraction voltage resulted in the removal of tails in phase space. Beam based steering in beamlines led to the reduction of beam emittance growth. Readjustment of the 100-MeV quadrupole transport resulted in the elimination of excessive beam envelope oscillations and removed significant parts of the beam halo at the target. Careful beam matching in the drift tube linac (DTL) provided high beam capture (75%) in the DTL. After improvements, beam losses in the 100-MeV beamline were reduced by an order of magnitude and reached the fractional level of 5 x10$^{-4}$.

## ISOTOPE PRODUCTION AT LANL

The Isotope Production Facility is a part of the Los Alamos accelerator facility. It has been in operation since 2004, producing isotopes for a variety of applications [1]. Rubidium chloride and gallium metal targets are irradiated by a 100-MeV proton beam with average proton beam current of 250 μA. Three target slots allow target irradiation to be optimized by energy range for a particular isotope. Strontium-82 produced at Los Alamos is used in hospitals nation- and world-wide for critical cardiac imaging while LANL-produced germanium-68 is used to calibrate Positron Emission Tomography (PET) equipment used for a broad spectrum of diagnostic procedures. Some isotopes produced, such as aluminum-26 and silicon-32, are unique to Los Alamos, while others, such as the burgeoning production of Ac-225, are available from multiple sites. Production beam time is around 3000 hours / year. Future plans include increase of average beam current up to 450 μA and operation with various beam energies of 41, 72 and 100 MeV.

The proton beam for isotope production is generated by a duoplasmatron proton ion source mounted to a 750-keV Cockroft-Walton accelerating column. The 750-keV Low-Energy Beam Transport line (LEBT) delivers beam to Drift Tube Linac (see Figs. 1, 2). After acceleration in the DTL, 100-MeV proton beam is transported in the Transition Region (TR) and in the IPF beamline, where it is incident to the IPF production target. Horizontal and vertical raster-steering magnets perform circular sweeping of the beam on the IPF target. Optimization of proton beam performance was required to insure uninterrupted beam delivery with high reliability and preparation to enhance of isotope production capacity.

## PROTON ION SOURCE

The current proton ion source is a duoplasmatron with a Pierce extraction geometry. The source was originally developed for production of 50-mA beam. Later, the source was modified for greater brightness with lower current (20 - 25 mA) [2]. Between 2006 - 2014, the LANSCE accelerator was in operation at 60 Hz pulse rate to prevent excessive cathode emission and ceramic cracking in 201.25 MHz amplifiers feeding the DTL. At that time, proton beam source was operated at 40 Hz x 625 μsec pulse length delivering 14 mA peak current beam. In 2014 the LANSCE accelerator facility returned to 120 Hz pulse rate operation. Presently the source delivers a proton beam with a current of 5 mA at 100 Hz x 625 μsec pulse length.

Normalized beam emittance extracted from ion source is estimated as [3]

$$\varepsilon_{rms} = \frac{R}{2} \sqrt{\frac{k_B T}{mc^2} + (\frac{qBR}{4mc})^2} \; , \qquad (1)$$

where $R$ is the aperture radius, $k_B = 1.38 \times 10^{-23}$ J/K is the Boltzman constant, $T$ is the plasma temperature, and $B$ is the magnetic field between intermediate electrode and anode. For typical source parameters $R = 3$ mm, $k_B T = 0.1$ eV, $B = 200$ G, the Eq. (1) gives estimation of rms normalized beam emittance as $\varepsilon_{rms} = 0.0017$ π·cm·mrad. Additional sources contributing to beam emittance are: irregularities in the plasma meniscus extraction surface, aberrations due to ion-source extraction optics, optical aberrations of the focusing elements of the Low Energy Beam Transport, non-linearity of the electric field created by the beam space charge, and beam fluctuations due to ion-source instability or power regulation [4]. The extraction of the beam from ion source is characterized by the beam perveance

$$P_b = \frac{I}{U_{ext}^{3/2}}, \qquad (2)$$

where $I$ is the beam current and $U_{ext}$ is the extraction voltage. Geometry of extraction region is characterized by the Child-Langmuir perveance:

$$P_o = \frac{4\sqrt{2}\pi}{9} \varepsilon_o \sqrt{\frac{q}{m}} S^2, \qquad (3)$$

where $S = R/d$ is the ratio of extraction radius to extraction gap (aspect ratio of extractor). The ratio of beam perveance to Child-Langmuir perveance

$$\eta = \frac{P_b}{P_o} = \frac{9}{\sqrt{2} S^2} \frac{I}{I_c} \left(\frac{mc^2}{qU_{ext}}\right)^{3/2}, \qquad (4)$$

where $I_c = 4\pi\varepsilon_o mc^3/q = 3.13 \times 10^7 A/Z$ [Amp] is the characterisctic current, is used to optimize extracted beam quality [4]. Recent study of LANSCE proton source [5] indicates that beam brightness is maximized for optimal value of matching parameter $\eta_{opt} = 0.52$. Based on obtained results, the dependence of extraction voltage on beam current for maximizing beam brightness can be expressed from Eq. (4) as

$$U_{ext} = \left(\frac{I}{P_o \eta_{opt}}\right)^{2/3}, \qquad (5)$$

where Child-Langmuir perveance for this source is $P_o = 9.16 \cdot 10^{-9} \, A/V^{3/2}$.

Historically, extraction voltage of $U_{ext} = 30$ kV was used to deliver beam with peak current 23 mA for operation of accelerator with average current of 1 mA. At the time of delivering 14 mA of proton beam for isotope production, the value of extraction voltage was selected to be $U_{ext} = 27$ kV. Transformation from 14 mA to 5 mA beam extraction resulted in appearance of long tails in beam emittance (see Fig. 3a). Expected value of extraction voltage from Eq. (5) at the beam current of 5 mA to maximize beam brightness is $U_{ext} = 10 \, kV$. Ion source is mounted to a 750-keV Cockroft-Walton accelerating column. First electrode of Cockroft-Walton column is served as an extraction electrode for ion source. For stable operation of accelerating column, total voltage of 750 kV is distributed among 16 electrodes. For this reason, the extraction voltage cannot be made less than 20 kV. For stable operation, it was selected to be $U_{ext} = 21$ kV. Reduction of the voltage from 27 kV to 21 kV resulted in significant suppression of beam tails (see Fig. 3b). Total beam emittance was reduced from 0.03 π·cm·mrad to 0.013 π·cm·mrad, while rms emittance dropped from 0.011 π·cm·mrad to 0.003 π·cm·mrad. Typical value of the normalized rms proton beam emittance after optimization is $\varepsilon_{rms} = 0.002 - 0.003$ π·cm·mrad. The ratio of total emitance to rms emittance ot the extracted beam is typically $\varepsilon_{total}/\varepsilon_{rms} = 5...7$, which indicates that the beam distribution is between "Water Bag" and parabolic distribution in 4D phase space [6].

## 750 KEV H$^+$ BEAM TRANSPORT

After extraction, proton beam is accelerated in 750 kV Cockcroft-Walton column followed by 11 m Low-Energy Beam Fransport (LEBT) line. The beamline consists of a quadrupole lattice, 81° and 9° bending magnets, RF prebuncher, main buncher, diagnostics and steering magnets. The beam transport lattice starts with six quadrupoles (TAQL1-TAQL4) before the emittance station TAEM1. The two quadrupole doublets TAQL5_1-2, TAQL6_1-2, and a 81° bending magnet match the beam in front of the pre-buncher creating a beam waist with transverse size of 0.4 cm. The quadruplet TAQL7_1-4 matches beam to the ground level deflector. The TAQL8_1-4 quadruplet creates a beam waist in front of the main buncher, with transverse beam size of 0.4 cm. The quadruplet TDQL4_1-4 matches the beam to drift tube linac. The Low Energy Beam Transport includes five steering magnets TASM1-5 and the common steering magnet TDSM1 shared with the H$^-$ beam. Slit-collector beam emittance measurements are performed using three stations: 1) TAEM1 (just after the Cockroft -Walton column), 2) TAEM2 after pre-buncher, and 3) TDEM1 (before the entrance to the DTL). Additionally, there are three harps (TAHP1-2, TDHP1) used for control of the beam profile. They are located close to the slit positions of each emittance station.

Focusing properties of proton LEBT are similar to that of H$^-$ beam transport [7]. Normalized acceptance of the channel can be estimated as

$$\varepsilon_{ch} \approx \beta\gamma \frac{\mu_o}{S(1+\upsilon_{max})^2} a^2, \qquad (6)$$

where $\beta\gamma$ is the particle momentum, $\mu_o$ is the phase advance of transverse oscillations at distance $S$, $a$ is the aperture of the channel, and $\upsilon_{max}$ is the envelope ripple factor. In the beamline, single particle performs one complete oscillation at the distance of 6 m. The aperture of the channel is $a \approx 2$ cm, and ripple factor is $\upsilon_{max} \approx 0.7$. Based on these numbers, the normalized acceptance of the channel is estimated as $\varepsilon_{ch} \approx 0.55$ π cm mrad, while more accurate numerical simulations provide value of $\varepsilon_{ch} = 0.4$ π cm mrad. Space charge significantly affect beam dynamics. Space charge factor of depression of transverse oscillations in beamline is [8]

$$\frac{\mu}{\mu_o} = \sqrt{1+b_o^2} - b_o = 0.41, \qquad (7)$$

where $b_o$ is the space charge parameter:

$$b_o = \frac{1}{(\beta\gamma)^2} \frac{I}{I_c} \frac{S}{\mu_o(4\varepsilon_{rms})} \approx 1. \qquad (8)$$

Emittance growth caused by space charge is estimated as

$$\frac{\varepsilon_f}{\varepsilon_i} = \sqrt{1+(\frac{\mu_o^2}{\mu^2}-1)(\frac{W_i-W_f}{W_o})}, \qquad (9)$$

where $(W_i-W_f)/W_o$ is the "free-energy" parameter due to space-charge induced beam intensity redistribution. Parameter $(W_i-W_f)/W_o$ has a value of 0.01126 for "Water Bag" distribution, and 0.0236 for parabolic beam distribution in 4D phase space. The estimated space charge emittance growth from Eq. (9) is between 3% - 6 %. In experiments, no noticeable emittance growth due to space charge at beam current of 5 mA was observed. However, space charge plays a significant role in beam matching to the structure.

## BEAM BASED ALIGNMENT

Operation of low-energy beam transport indicated that the beam emittance growth is mostly determined by misalignment of beamline components. Recent alignment measurements in the accelerator facility performed by the LANSCE Mechanical Team indicated that most of components have mutual displacement within ±1 mm (see Fig. 4). The well-established procedure for beam-based alignment utilizes multiple measurements of the beam centroid position and the solution of the resultant matrix to determine the offsets of the magnets, beam position monitors (BPM) and beam centroids [9].

This beam steering procedure was applied subsequently to portions of the beamline containing combinations of quadrupoles preceded by a steering magnet and a harp to monitor the beam centroid (see Fig. 5). Because only single steering magnet is used for beam-based alignment in a group of quadrupoles, the procedure is applied to steer the beam in a selected (usually central) quadrupole with subsequent verification of steering effect in the whole collection of quadrupoles. Initial setup required all magnets to be off in the region being evaluated. To determine beam offset and beam angle at the entrance of a selected quadrupole, the value of this quadrupole field was varied from zero up to a point were visible displacement of the beam could be seen on the measurement device. Dynamics of the beam centroid at the entrance of quadrupole on to the measurement device is determined by the product of the quadrupole and drift space matrices:

$$\begin{pmatrix} a_{11} & a_{12} \\ a_{21} & a_{22} \end{pmatrix} = \begin{pmatrix} 1 & L_i \\ 0 & 1 \end{pmatrix} \begin{pmatrix} Q_{11} & Q_{12} \\ Q_{21} & Q_{22} \end{pmatrix}, \tag{10}$$

where $Q_{ij}$ are elements of matrix of a quadrupole (focusing or defocusing), and $L_i$ is the distance from selected quadrupole to the measurement device (see Fig. 5). Taking three measurements at three different values of the quadrupole field, one gets the system of equations:

$$x_1 = a_{11}(1)x_o + a_{12}(1)x_o' + \delta x, \tag{11}$$

$$x_2 = a_{11}(2)x_o + a_{12}(2)x_o' + \delta x, \tag{12}$$

$$x_3 = a_{11}(3)x_o + a_{12}(3)x_o' + \delta x, \tag{13}$$

where observed values of beam centroid displacements $x_1$, $x_2$, $x_3$ depend on the offset of the measurement device with respect to quadrupole axis, $\delta x$. Taking the difference between measurements $\Delta x_1 = x_2 - x_1$ and $\Delta x_2 = x_3 - x_2$,

$$\Delta x_1 = [a_{11}(2) - a_{11}(1)]x_o + [a_{12}(2) - a_{12}(1)]x_o', \tag{14}$$

$$\Delta x_2 = [a_{11}(3) - a_{11}(2)]x_o + [a_{12}(3) - a_{12}(2)]x_o', \tag{15}$$

one gets the system of equations to determine beam offset and centroid angle, $x_o$, $x_o'$, which are independent on the measurement device offset:

$$x_o' = \frac{\Delta x_2[a_{11}(2)-a_{11}(1)] - \Delta x_1[a_{11}(3)-a_{11}(2)]}{[a_{11}(2)-a_{11}(1)][a_{12}(3)-a_{12}(2)] - [a_{11}(3)-a_{11}(2)][a_{12}(2)-a_{12}(1)]}, \tag{16}$$

$$x_o = \frac{\Delta x_1 - [a_{12}(2) - a_{12}(1)]x_o'}{a_{11}(2) - a_{11}(1)} . \qquad (17)$$

Correction of beam displacement at a quadrupole is achieved through trajectory kick applied from steering magnet. Calibration of steering magnets is performed through two measurements of the beam centroid at the measurement device $x_{1s}$, $x_{2s}$, while changing the current of the steering magnet $\Delta I_{sm}$, and keeping all subsequent quadrupoles off. Strength of steering magnet is determined by

$$k = \frac{(x_{2s} - x_{1s})}{L_s \cdot \Delta I_{sm}} , \qquad (18)$$

where $L_s$ is the drift length from the steering magnet to measurement device (see Fig. 5). The required steering kick, $\Delta x'$, is estimated from condition of compensation of the beam offset at a selected quadrupole:

$$\Delta x' = -\frac{x_o}{S_i} , \qquad (19)$$

where $S_i$ is the distance from the steering magnet to selected quadrupole. This offset is affected by beam misalignments in preceding quadrupoles. The change of steering magnet current to get a desired kick is estimated by $\Delta I_{sm} = \Delta x'/k$. The value of $\Delta x'$ has to be eventually corrected to provide minimum beam offsets in a group of quadrupoles (see Fig. 6).

The final step is verification of beam offset in each quadrupole. After selection of steering magnet kick, the field in each quadrupole was varied from zero until displacement is seen. All downstream quadrupoles were kept off. This process began with the quadrupole nearest to the steering magnet and was repeated until the last quadruple before measurement device. Beam offset in *i*-th quadrupole is given by

$$\Delta_i = \frac{(x_{Quad \neq 0} - x_{Quad=0})}{D_i L_i \Delta G_i} \left( \frac{mc\beta\gamma}{q} \right) , \qquad (20)$$

where $\Delta x_i = x_{Quad \neq 0} - x_{Quad=0}$ is the variation of beam centroid at measurement device, $\Delta G_i$ is the change in quadrupole gradient and $D_i$ is the quadrupole length. Equation (20) is the result of thin-lens approximation of matrix equation (10), which led to Eq. (17). The "zero quadrupole steering" corresponds to the minimal value of rms variation of the beam centroid position among all *N* quadrupoles:

$$\Delta = \sqrt{\frac{1}{N} \sum_{i=1}^{N} \Delta_i^2} . \qquad (21)$$

Described analysis is applied to beam centroid, which is weakly affected by beam space charge. Phase advance of coherent oscillations of the beam centroid, $\mu_{coh}$, in presence of space charge, is described by [6]

$$\mu_{coh}^2 = \mu_o^2 - \left(\frac{S}{a}\right)^2 \frac{2I}{I_c(\beta\gamma)^3} . \qquad (22)$$

Substitution of LEBT parameters into Eq. (22) determines the value of phase shift of coherent oscillation as

$$\frac{\mu_{coh}}{\mu_o} = 0.996, \qquad (23)$$

which is practically the same as that for single oscillation particle without space charge forces.

## APPLICATION OF BEAM BASED ALIGNMENT TO THE PROTON LEBT

The described procedure was used to provide beam based steering of the low-energy proton beam transport. The first six quadrupoles (TAQL1–TAQL4) after the 750-keV Cockroft-Walton accelerating column are not preceded by a steering magnet. Variation of field gradients in these quadrupoles did not result in any significant variation of the beam centroid. Centering of the beam began with the quadrupole doublet TAQL5_1, TAQL5_2 in front of the 81° bending magnet.

Beam based alignment of quadruple TAQL5_1 started with determination of beam offset in TAQL5_1. Quadrupoles TAQL5_1, TAQL5_2, TAQL6_1, TAQL6_2 were powered off. Beam centroid variation was observed at harp TAHP02, while gradient at TAQL5_1 was varied from zero to 46 Gauss/cm. It resulted in change of beam centroid at harp $\Delta x = 0.147$ cm, $\Delta y = -0.153$ cm. Change of matrix elements $\Delta a_{11\_x} = 0.3098$, $\Delta a_{11\_y} = -0.3736$, $\Delta a_{12\_x} = 1.63 \times 10^{-3}$ cm/mrad, $\Delta a_{12\_y} = -2.03 \times 10^{-3}$ cm/mrad were determined using TRACE code [10]. Incoming beam slopes $x_o' = 1.717$ mrad, $y_o' = 2.366$ mrad, were determined from beam emittance scans. The values of the beam offset at the entrance of TAQL5_1 using equation (17) are:

$$x_o = \frac{\Delta x - \Delta a_{12\_x} \cdot x_o'}{\Delta a_{11\_x}} = 0.465 \, cm, \qquad (24)$$

$$y_o = \frac{\Delta y - \Delta a_{12\_y} \cdot y_o'}{\Delta a_{11\_y}} = 0.396 \, cm. \qquad (25)$$

Calibration of steering magnets provided the value of the steering magnets strength $k = 1.2$ mrad/A. The distance from steering magnet TASM01 to quadrupole TAQL5_1 was $S_i = 163.251$ cm. The required kick from steering magnet TASM01 to compensate beam offset at quadrupole doublet TAQL5_1/TAQL5_2 was:

$$\Delta x' = -\frac{x_o}{S_i} = -2.85 \text{ mrad}, \qquad \Delta y' = -\frac{y_o}{S_i} = -2.42 \text{ mrad}. \qquad (26)$$

Minimization of beam offset in next quadrupole doublet TAQL6_1, TAQL6_2 was performed with variation of the field for bending magnet TABM01 and the subsequent determination of beam offset in each quadrupole. The value of vertical steering TASM01_V kick was corrected to - 3.6 mrad to provide minimum beam offset in quadruplet TAQL5 - TAQL6.

This method was applied in the rest of beam transport to determine the required steering. Steering magnet TASM3 was used to align beam in quadruplet TAQL7_1-4. Steering TASM4 was used to steer beam in quadruplet TAQM8. After steering, determined setup of low-energy beam transport provided significantly smaller beam emittance growth. Table 1 illustrates beam emittance growth along LEBT before and after beam alignment. Figure 7 illustrated beam emitance plots at the end of LEBT. As seen, emittance growth, and final LEBT beam emittance, were reduced by a factor of two.

## DYNAMICS OF ACCELERATED BEAM

After transport, beam enters 201.25 MHz Drift Tube Linac. The DTL consists of 4 tanks with energies 5 MeV, 41 MeV, 73 MeV, and 100 MeV, respectively. Originally designed for operation with synchronous phase of -26°, the linac was historically retuned for -32°, -23°, -22°, -32° synchronous phase setup with field amplitude of 98%, 96%, 94%, 98% of nominal values to minimize beam spill. Proton beam is preliminary bunched in LEBTs in 2-buncher systems. Beam are captured in DTL with efficiency of 75% - 80% , so 20% - 25% of the beam is lost in Tank 1 due to insufficient bunching. Subsequent beam losses of 0.1% - 1% in DTL are determined by un-captured particles, and by expansion of phase space volume occupied by the beam.

Proton beam experiences additional emittance growth in DTL. Normalized rms emittance after acceleration in DTL is 0.02 π cm mrad. Normalized acceptance of DTL is changing from 1 π·cm·mrad in the beginning of accelerator to 2 π·cm·mrad at the end. Total beam emittance can be approximated by the value of $6\varepsilon_{rms}$. Ratio of normalized acceptance of DTL accelerator to $6\varepsilon_{rms}$ beam emittance is varied from 40 in the beginning to 16 at the end of DTL. Beam emittance growth in DTL is mostly due to transverse-longitudinal coupling in RF field and misalignment of accelerator structure.

After the DTL, 100-MeV protons enter the Transition Region (TR) and continue on to the IPF beamline. Transition Region and Isotope Production Facility beamlines include four bending magnets (total 45° bend), nine quadrupole magnets, six steering magnets, four current monitors, eight beam position monitors (BPMs), a wire scanner and a harp for beam profile control (see Figs. 8, 9). Operation of TR and IPF beamlines include BPM control of the beam centroid, correction of beam position at the target and control of beam losses at the Activation Protection (AP) devices. Each AP device is a one-pint can, consisting of a photomultiplier tube immersed in scintillator fluid. The AP detectors integrate the signals and shut off the beam if the losses exceed the specified limit. They are calibrated with 6-μA point beam losses at the energy of 100 MeV. Most of beam losses are observed around IPAP03 (see Fig. 8). Typical beam losses in IPF beamline are characterized by summed AP device readings of 15% - 20%, which is equivalent to 1-μA beam losses, or relative beam losses of $4 \times 10^{-3}$.

Series of beam development experiments were undertaken to reduce beam losses. Analysis of beam dynamics, using 100-MeV beam emittance scan, indicated that beam envelopes had excessive variation (see Fig. 10a). New quadrupole setup for quadrupoles TRQM01-02, IPQM01-07 provided more relaxed beam envelopes during transport (see Fig. 10b). Wire scans at IPWS01 confirmed improvements of beam dynamics (see Fig. 11).

Additional improvement of beam quality was achieved with realignment of the IPF beamline. Laser tracking of beam elements, performed by the LANSCE Mechanical Team, indicated noticeable displacements of the magnets (see Fig. 12). A combination of the steering and bending magnets were adjusted to center the beam through the sequence of quadrupoles. The procedure used was similar to that applied to Low Energy Beam Transport. Steering magnets IPHM/VM01-04 were adjusted to minimize beam displacements in quadrupoles IPQM1-7. Beam position monitors IPPM01-09 indicated noticeable alignment of beam centroid trajectory (see Fig. 13). As a result of improved beam matching, the beam losses were reduced significantly from $4 \times 10^{-3}$ to $5 \times 10^{-4}$, which corresponds to a reading at IPAP03 of 2%, or 0.12 μA (see Table 2). Overall optimization of proton beam resulted in removing of excessive beam halo at the Isotope Production target (see Fig. 14) and provide clean beam sweeping of the target using circular rastering (see Fig. 15).

## SUMMARY

A series of beam development experiments were undertaken to reduce emittance growth and beam losses in LANSCE Isotope Production Facility. A beam based steering procedure in Low Energy Beam Transport and in IPF beamline was implemented to minimize emittance growth. It included the determination of beam offset and beam angle upon entering a group of quadrupoles, with subsequent correction of beam centroid trajectory to minimize beam offset in a series of quadrupoles. Application of this procedure resulted in significant reduction of emittance growth in Low Energy Beam Transport. Analysis of dynamics of 100-MeV beam in IPF beamline indicated that beam envelopes had excessive variation, which was corrected by quadupole setup. Additional improvement of beam quality was achieved by beam based alignment in the IPF beamline. A combination of the steering and bending magnets were adjusted to center the beam through the sequence of IPF quadrupoles. As a result of improved beam matching and steering, the beam losses were reduced from $4 \times 10^{-3}$ to $5 \times 10^{-4}$.

## ACKNOWLEDGEMENTS

Author is indebted to LANSCE Operation Team, Radioisotope Production Team, Mechanical Team for invaluable assistance, and to Charles Taylor for careful reading of manuscript and providing multiple useful comments.

## REFERENCES

[1] K. F. Johnson et al, Proceedings of EPAC 2004, Lucerne, Switzerland, p.2816.
[2] R.R Stevens, J.R. McConnell, E.P. Chamberlin, R.W. Hamm, and R. L. York, Proceedings of the 1979 Linear Accelerator Conference, Montauk, USA, September 10-14, 1979, p. 465.
[3] I. Ben-Zvi, Proceedings of the 1993 Particle Accelerator Conference, Washington, USA, May 17-20, p. 2962.
[4] C.Lejeune, "Extraction of High-Intensity ion Beams from Plasma Sources: Theoretical and Experimental Treatments", In: Applied Charged Particle Optics. Ed. By A.Septier, Part C, Academic Press, 1983.
[5] Y.Batygin, I.N. Draganic, C.M. Fortgang, Review of Scientific Instruments, 85, 103301 (2014).
[6] M.Reiser, Theory and Design of Charged Particle Beams, John Wiley & Sons, 1994.
[7] Y.K. Batygin, NIM-A 904 (2018), 64-73.
[8] I.M. Kapchinsky, "Theory of Resonance Linear Accelerators", Harwood, 1985.
[9] C.E. Adolpsen et al, SLAC-PUB-4902 (1989).
[10] K.R. Crandall and D.P. Rusthoi, "Documentation for TRACE: An Interactive Beam-Transport Code", LA-10235-MS.

Table 1. Normalized rms beam emittance in LEBT (π cm mrad).

| Emittance Station | Before Beam Alignment (Horizontal/Vertical) | After Beam Alignment (Horizontal/Vertical) |
|---|---|---|
| TAEM1 | 0.003 / 0.002 | 0.003 / 0.002 |
| TAEM2 | 0.005 / 0.006 | 0.002 / 0.002 |
| TDEM1 | 0.006 / 0.007 | 0.004 / 0.003 |

Table 2. Activation protection devices reading before and after IPF beamline retuning.

| Activation Protection Device | Reading Before Retuning (%) | Reading After Retuning (%) |
|---|---|---|
| IPAP01 | 1 | 0 |
| IPAP02 | 0 | 0 |
| IPAP03 | 14 | 2 |
| IPAP04 | 0 | 0 |
| IPAP05 | 1 | 0 |
| IPAP06 | 2 | 0 |

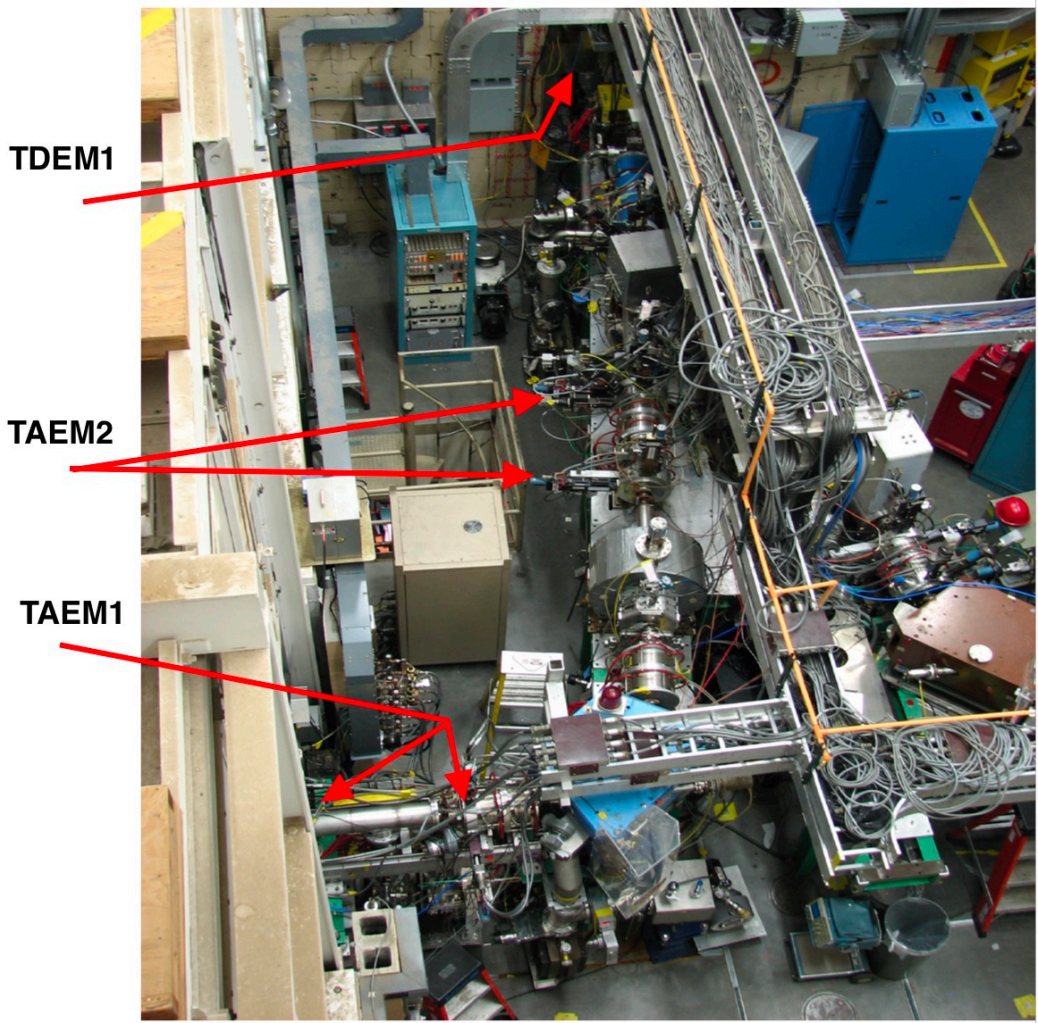

Figure 1: Layout of 750-keV proton Low Energy Beam Transport of LANSCE linear accelerator.

Figure 2: Focusing structure of 750-keV proton LEBT.

(a)

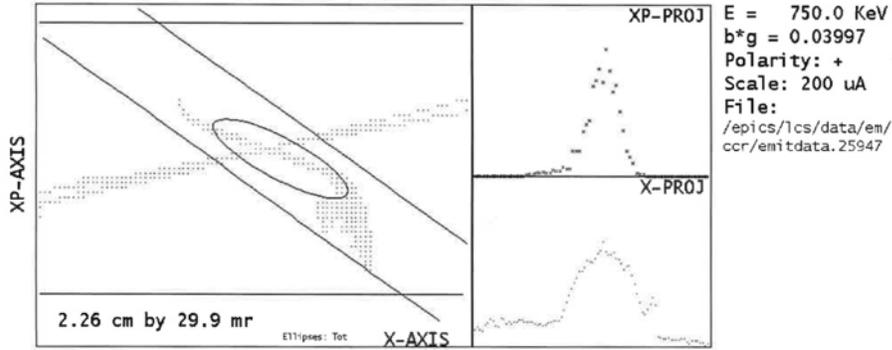

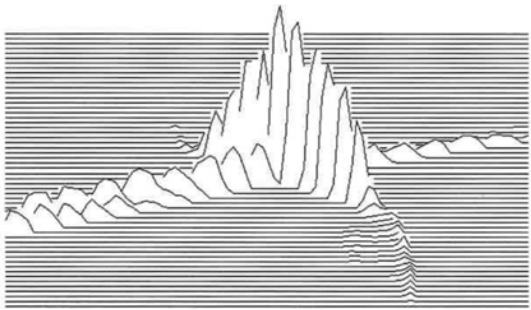

(b)

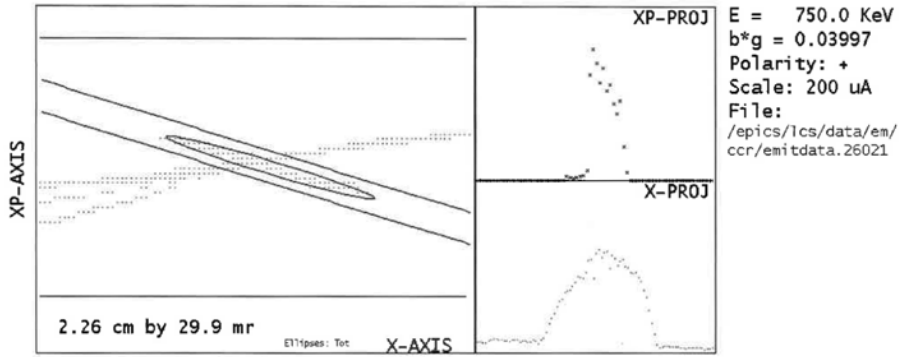

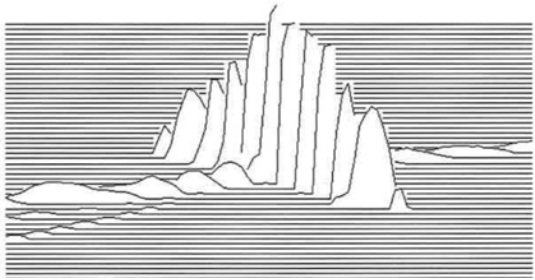

Figure 3: Emittance of the beam extracted from a proton ion source: (a) before and (b) after source adjustments. Beam distribution contains additional $H_2^+/H_3^+$ components.

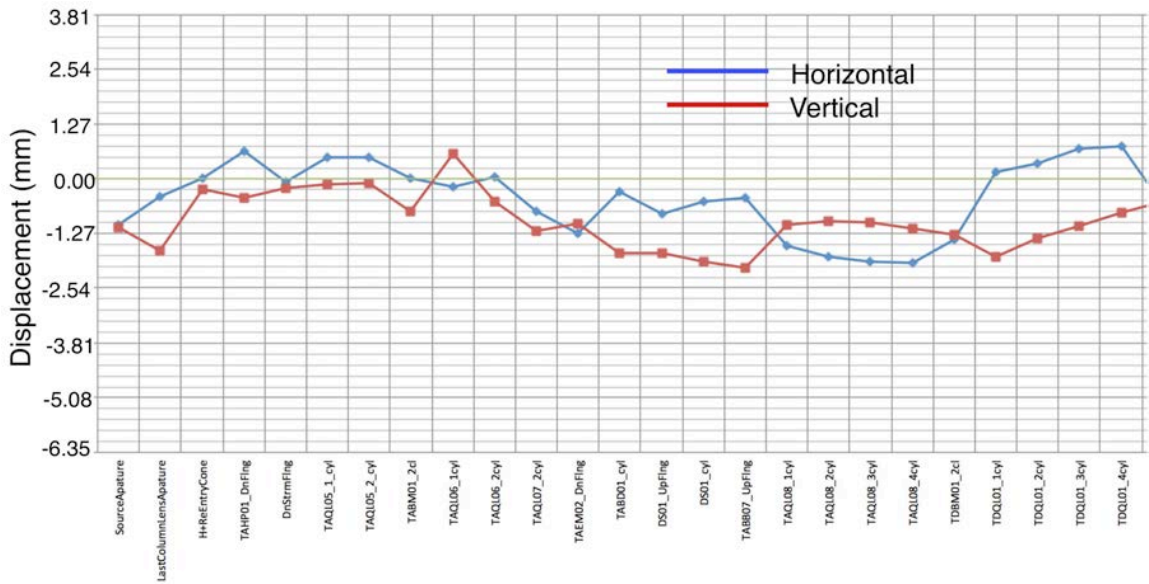

Figure 4: Measured misalignment of proton LEBT beamline elements.

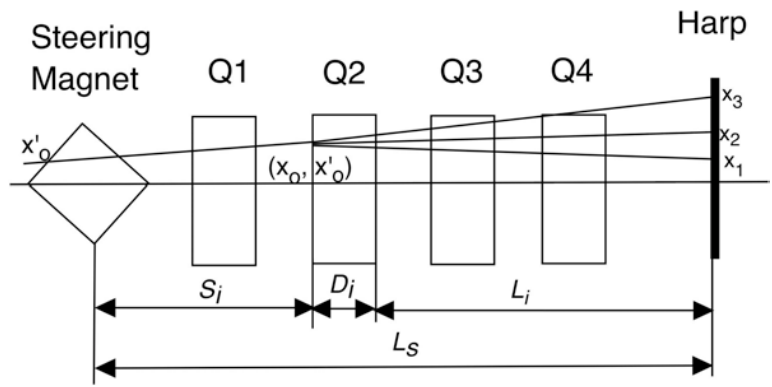

Figure 5: Determination of beam offset and beam angle at a quadrupole.

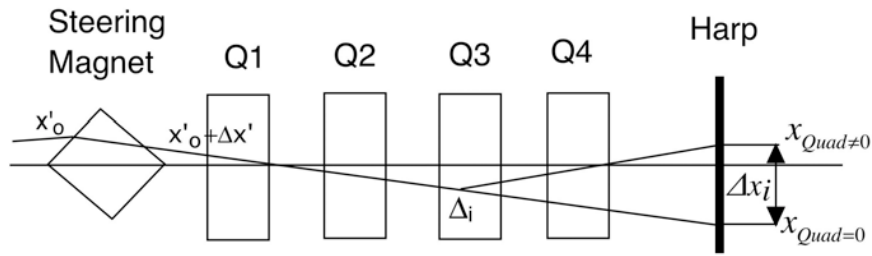

Figure 6: Minimization of beam offset in a sequence of quadrupoles.

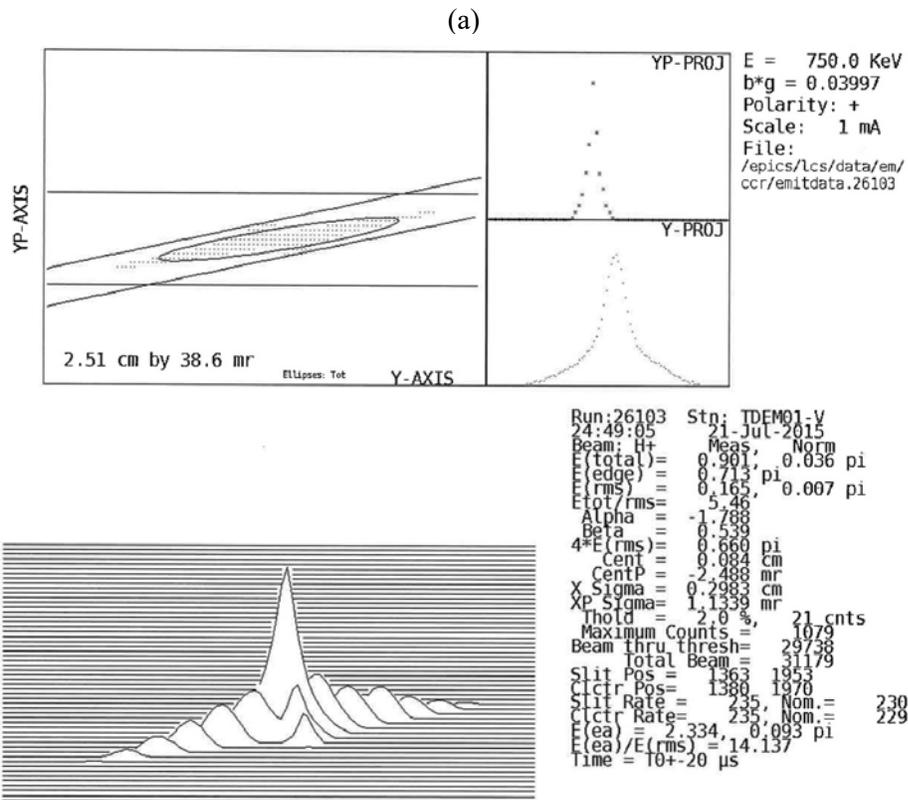

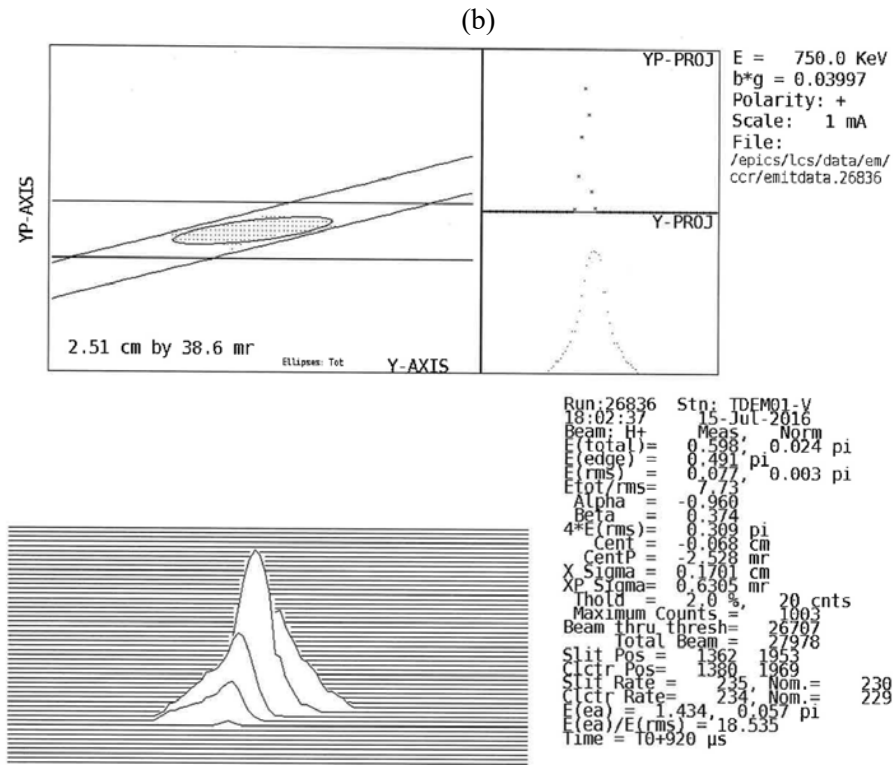

Figure 7: Emittance of the beam at the exit of LEBT: (a) before and (b) after beam based alignment.

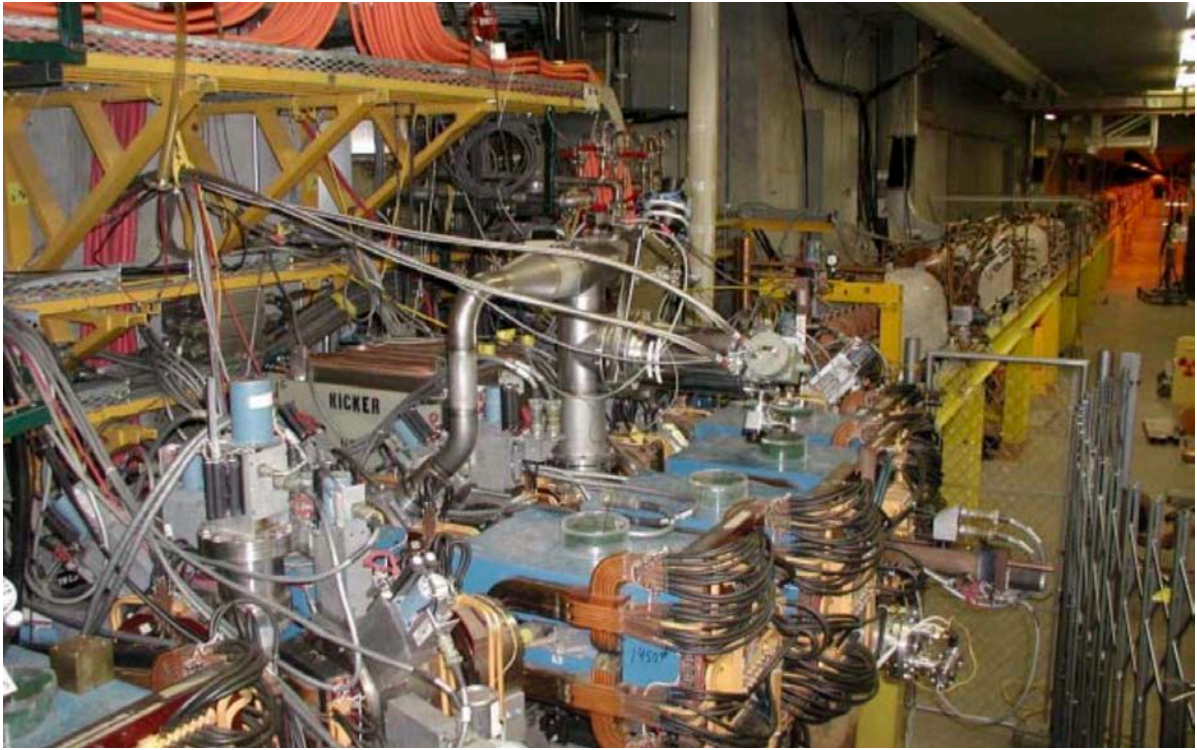

(a)

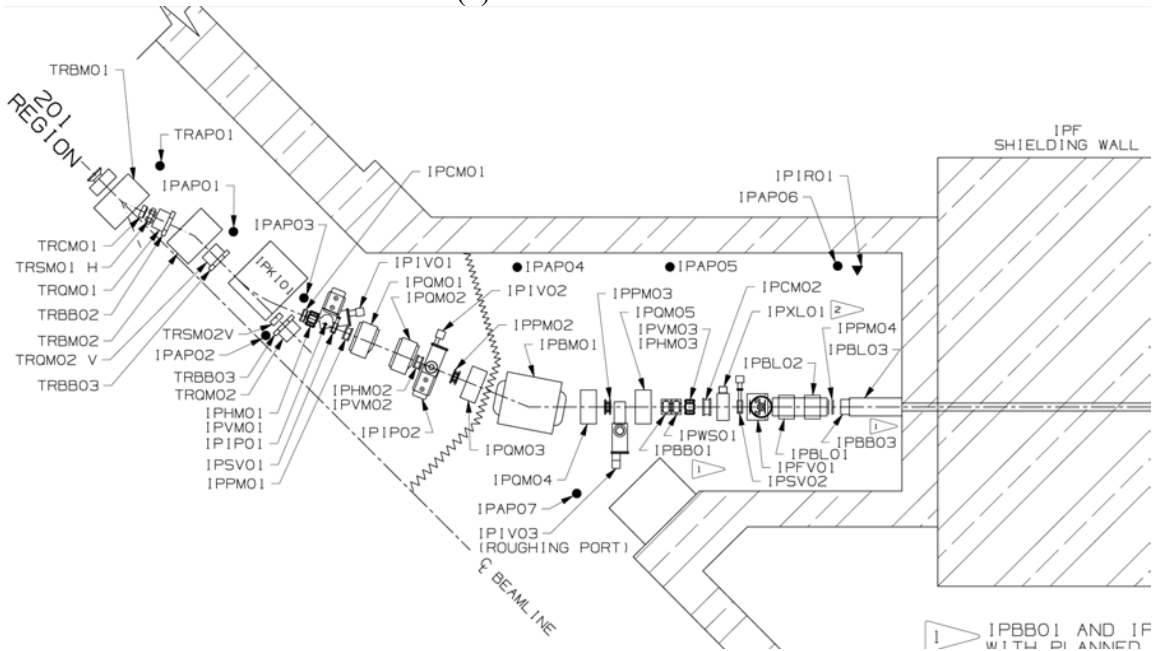

(b)

Figure 8: Transition Region from Drift Tube Linac to Isotope Production Beamline: (a) overview, (b) layout.

(a)

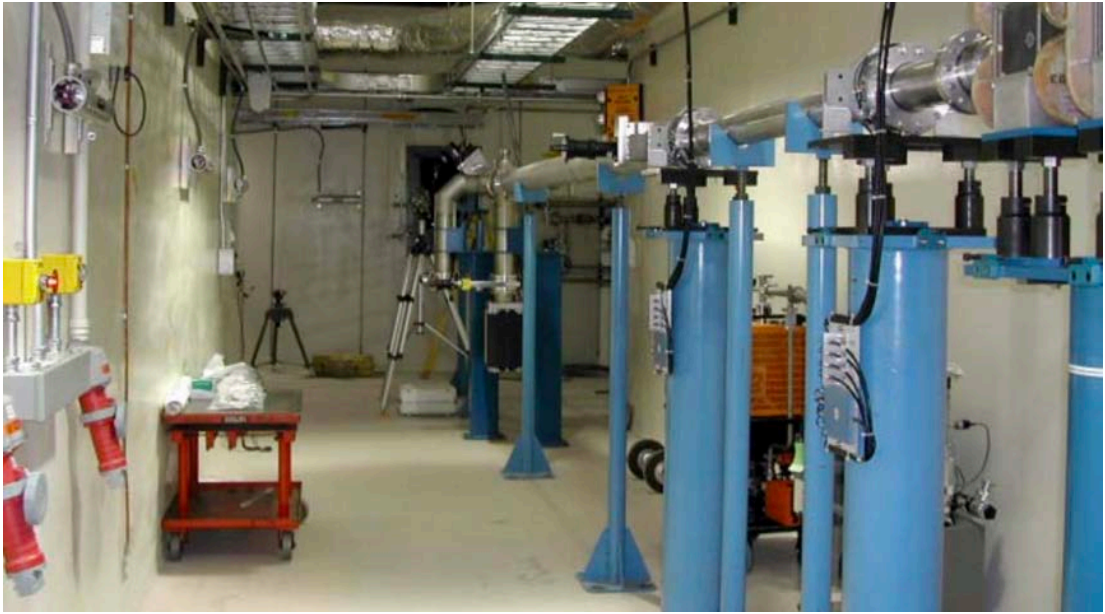

(b)

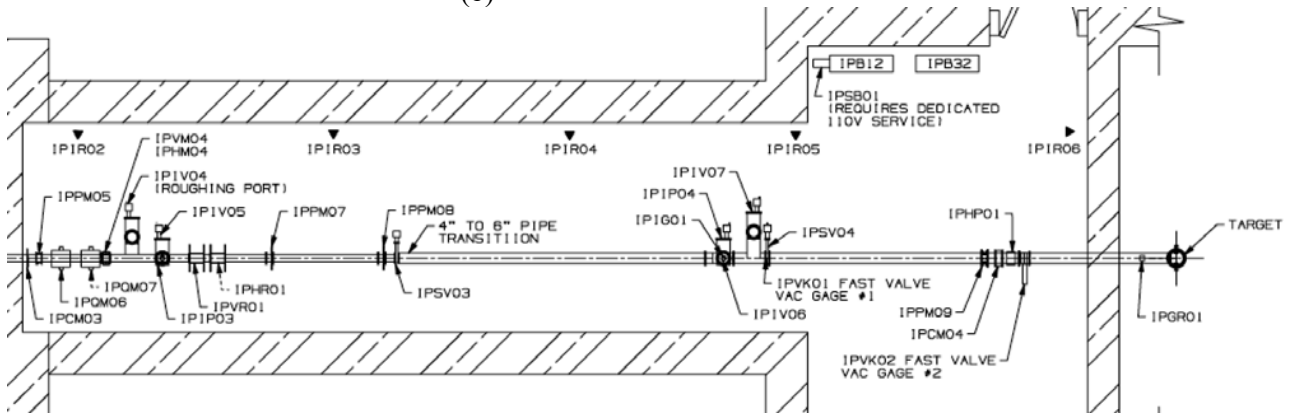

Figure 9: Isotope Production Facility beamline tunnel: (a) overview, (b) layout.

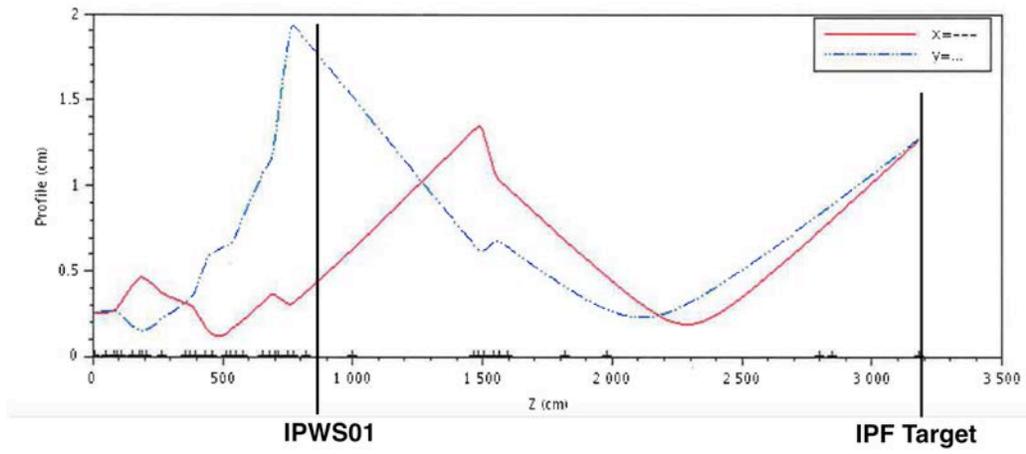

(a)

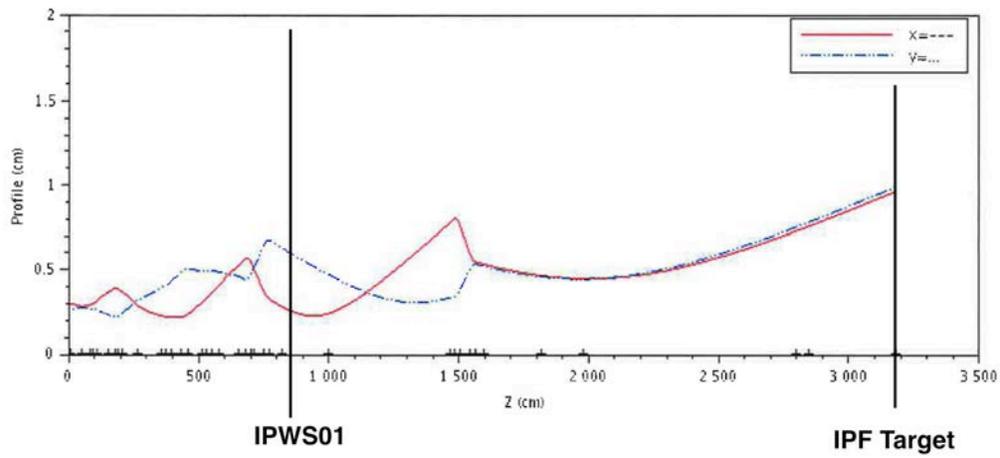

(b)

Figure 10: Beam envelopes (a) before and (b) after retuning of IPF beamline.

(a)

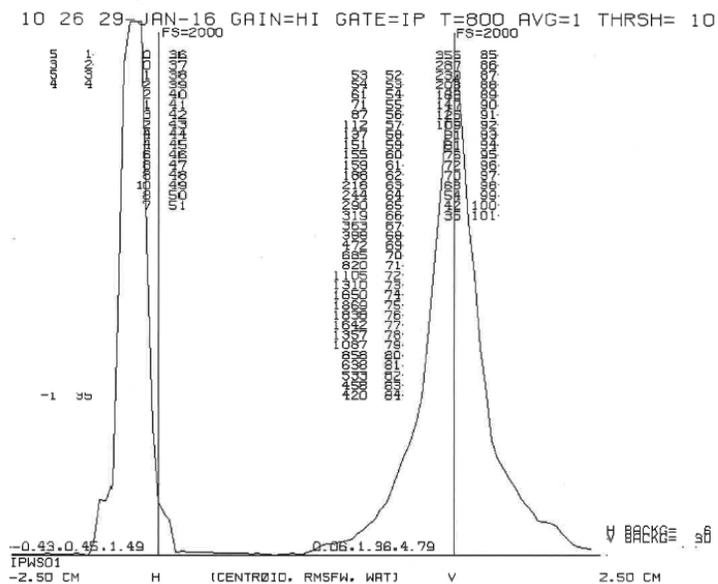

(b)

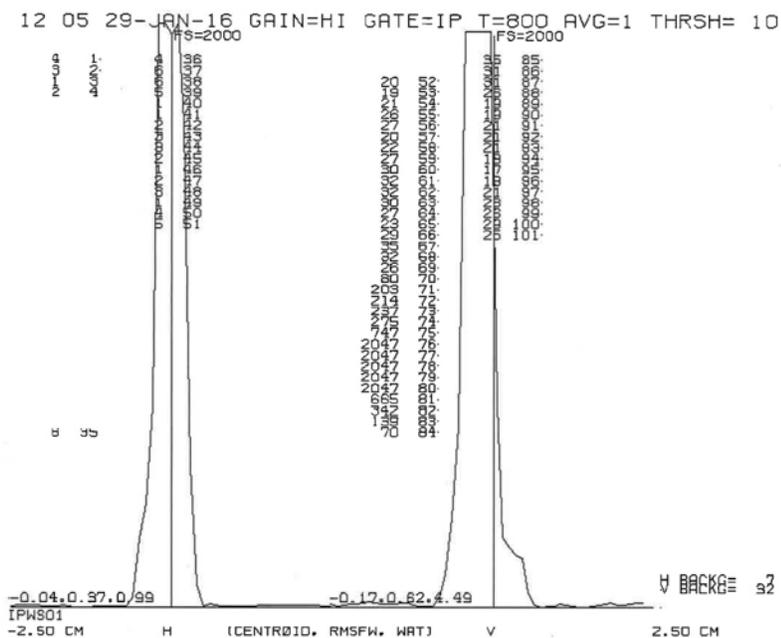

Figure 11: Beam profiles at IPWS01 wire scanner (a) before and (b) after retuning of IPF beamline.

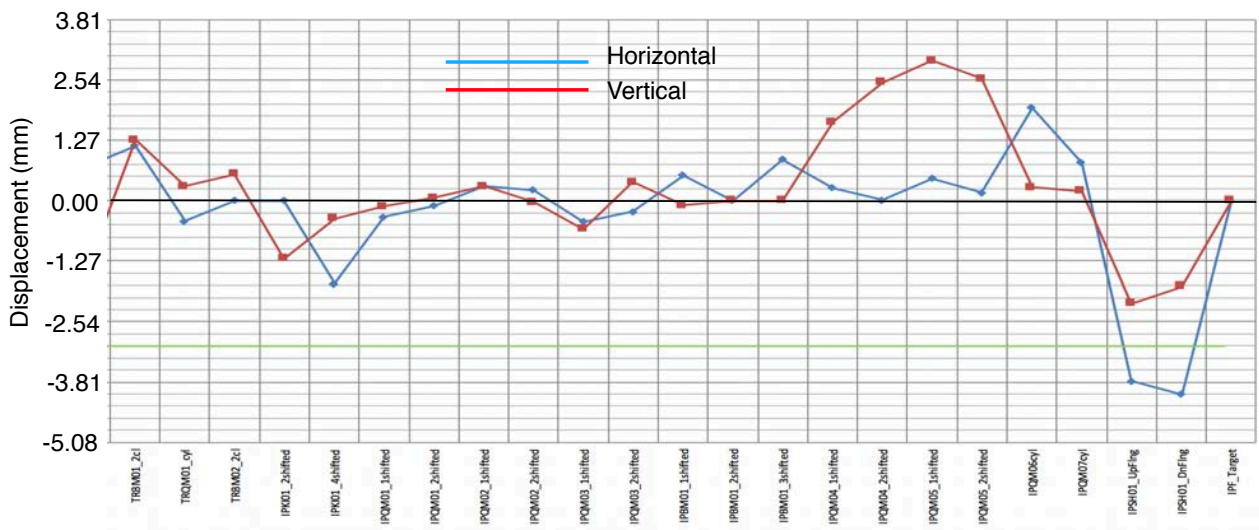

Figure 12: Misalignment of IPF beamline elements.

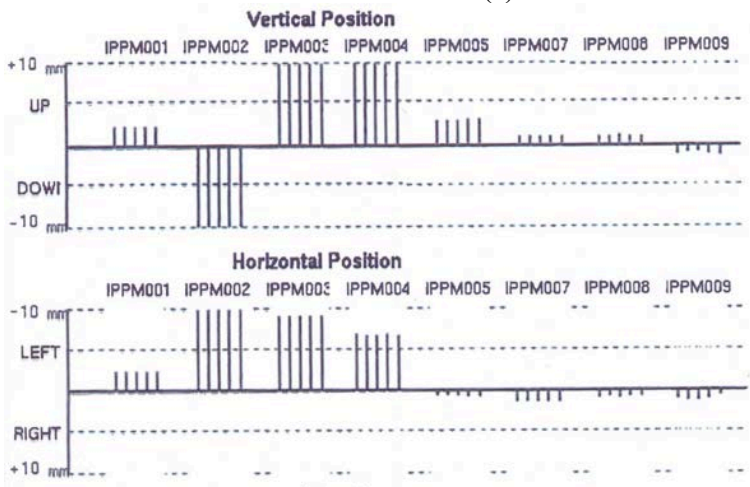

(a)

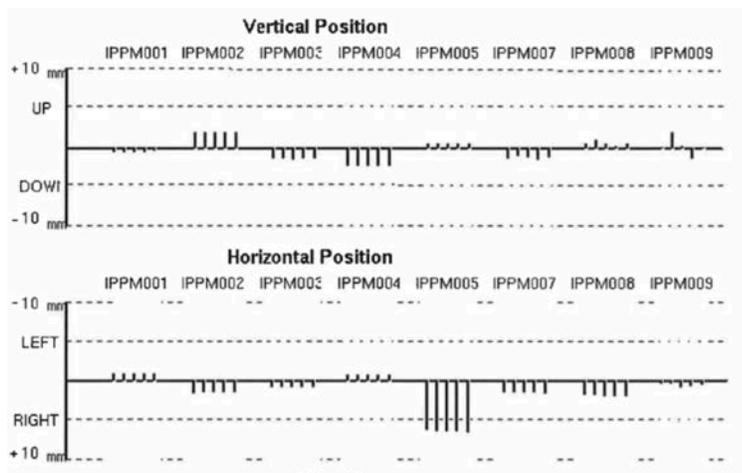

(b)

Figure 13: Beam position monitors bar graph (a) before and (b) after IPF beam based alignment.

(a)

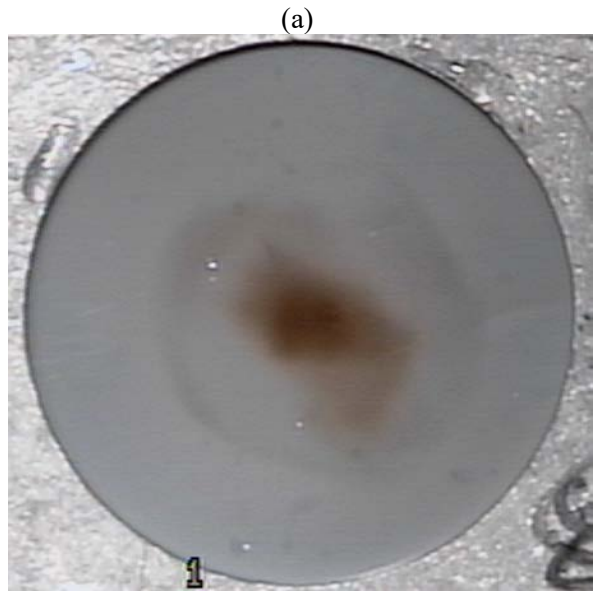

(b)

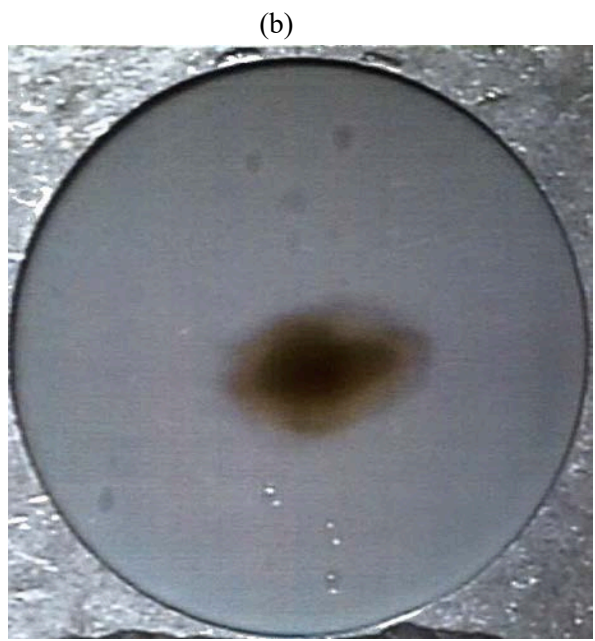

Figure 14. Unrastered beam at the target (a) before and (b) after retuning.

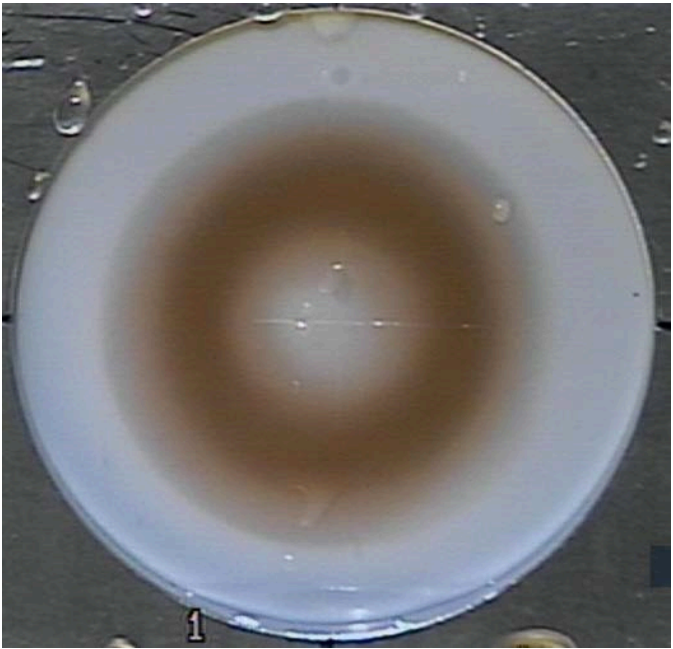

Figure 15: Rastered 100 MeV proton beam at Isotope Production Target.